\documentclass[a4paper,fleqn]{cas-dc}

\usepackage[numbers,super,sort&compress]{natbib}
\usepackage{graphicx}
\usepackage{subcaption}
\usepackage{amsmath}
\usepackage{placeins}

\def\tsc#1{\csdef{#1}{\textsc{\lowercase{#1}}\xspace}}
\tsc{WGM}
\tsc{QE}

\begin{document}
\let\WriteBookmarks\relax
\def\floatpagepagefraction{1}
\def\textpagefraction{.001}

\shorttitle{}    

\shortauthors{}  

\title [mode = title]{Popularity-driven interactions in multilayer bounded-confidence opinion dynamics}  

\author[1]{Senne De Bruyn}
\ead{senne.debruyn@ugent.be}

\author[2,1]{Luis E. C. Rocha}
\ead{luis.rocha@ugent.be}

\affiliation[1]{organization={Department of Physics and Astronomy, Ghent University}, country={Belgium}}
\affiliation[2]{organization={Department of Economics, Ghent University}, country={Belgium}}


\begin{abstract}
Social media users interact with their own social networks but also with platform-wide content, where a small number of highly popular users may receive a disproportionate share of attention. In this paper, we study how those interactions outside the social networks and user popularity regulate the formation of opinion clusters. We design a bounded-confidence opinion dynamics model on a multilayer network consisting of a social network coupled to a fully connected explore layer of potential interaction contacts. We show that uniform sampling on the explore layer increases mixing across social neighbourhoods and reduces opinion diversity while also weakening the influence of the underlying social network. We then introduce a popularity-driven interaction mechanism in which contact selection is biased by a time-evolving popularity score, so that frequently selected users become increasingly likely to be re-selected. The resulting concentration of explore-layer interactions on a small number of highly popular users reduces the homogenising effect of the explore layer, increasing opinion diversity. The resulting changes relative to a single-layer system depend on the underlying social network structure. As the popularity inequality increases, a small number of highly popular users can act as opinion attractors and sustain polarisation in regimes where the corresponding model without popularity heterogeneity converges to consensus.
\end{abstract}

\maketitle

\section{Introduction}\label{sec:intro}

On social media platforms, users are exposed to a variety of opinions on different topics and in different ways. On those platforms, users switch between different information feeds, where they can interact in different ways with friends, influencers, other like-minded users, or advertisements. Interactions have traditionally been constrained to a user-selected friendship network, but in more modern platforms, mechanisms of exploration have been introduced to allow users to interact with users, not necessarily like-minded, beyond their direct social network through recommendation, discovery, and advertising mechanisms~\citep{Bakshy2015}. Furthermore, users have been increasingly exposed to posts chosen and ranked via algorithmic personalisation based on user similarity, past activity, popularity scores, and platform rules~\citep{PerraRocha2019,Huszar2022}. Opinion dynamics models have been used to study the effects of these interactions and activity patterns on the evolution of the user opinions and on the emergence of opinion consensus and polarisation~\citep{Galam2005Heterogeneous, BoundedConfidenceSurvey2024, OpinionDynamicsRMP2026}.

Bounded-confidence models form an important class of opinion dynamics models in which interactions depend on the opinion-distance between individuals~\citep{BoundedConfidenceSurvey2024}. One of the most well-known continuous opinion dynamics models is the Deffuant-Weisbuch model~\citep{Weisbuch2001InteractingAA, DeffuantEtAl2000,weisbuch2002}. Different extensions of bounded-confidence models including heterogeneous behaviours have been studied. It has been shown, for example, that heterogeneity in stubbornness~\citep{GalamJacobs2007,WEISBUCH2005555,Lorenz_2009} can pull other users' opinions towards them, affecting opinion diversity or creating extremes. Susceptibility heterogeneity of social media users and the presence of some weakly susceptible agents can also affect opinion diversity and create extreme opinions~\citep{laguna2003minoritiesmodelopinionformation,DeffuantAWF02}. It can also affect the rate at which opinions change~\citep{SHANG20143766}. Heterogeneity in the number of friends and the structure of the social network can affect opinion formation through features such as social clustering, highly connected users, and their interplay with personalisation, which may reinforce the formation of echo chambers~\citep{Meng_2018,GIRALDEZCRU20221256,Botte2022}. These models assume interactions via a single social layer. However, in social media, information can be fed via different layers and this can be studied by modelling the system as a multilayer social network influencing opinion formation~\citep{multipliZ,peralta2022opiniondynamicssocialnetworks}. Each layer may correspond to one mechanism of interaction, e.g., direct friendship interactions or platform-wide content recommendations. Previous research has shown that multiplexity may hinder consensus formation~\citep{multipliZ,antonopoulos2018opinionformationmultiplexnetworks}. For example, on a 2-layer scale-free network, the coupling of both layers can lead to more or less fragmentation than the single-layer version~\citep{XuanNguyen2018OpinionFO}.

Social interactions are also dynamic, with users forming and losing friends as time goes on, and receiving tailored information selected by algorithms. These temporal evolutions can be included in opinion dynamics models by giving the users the possibility of making and breaking social ties. These rewiring mechanisms can be random~\citep{Kozma_2008}, homophilic~\citep{WU2024129835,Maia_2021,kan2023}, or heterophilic~\citep{WU2024129835}. Rewiring can affect opinion diversity and consensus formation~\citep{Kozma_2008,kan2023}, and lead to the formation of echo chambers~\citep{Sasahara_2020}. Temporal evolution can also be included by considering time-evolving attributes of the users~\citep{Alessandretti2017Attractiveness,Steinbacher2024PreferenceMatching}, such as stubbornness~\citep{li2024boundedconfidencemodelsopiniondynamics,Lorenz_2009} or susceptibility. The timescales and heterogeneity of user-to-user interactions have been shown to also affect opinion formation~\citep{Holme_2012,Holme_2015,zarei2023fastmultipartisanburstscommunication,li2023boundedconfidencemodelopiniondynamics}. Some opinion dynamics models have also analysed the addition of highly influential agents, such as mass media or influencers. These agents are, under some circumstances, able to stir the opinions of other agents towards them via social pressure or nudging, thereby affecting the group opinion landscape~\citep{helfmann-2023,zhao-2015,pineda-2014}. In situations where individuals are more subjective to these highly influential agents, the whole population becomes subjective to the external influence~\citep{pansanella-2023}. Additionally, highly visible agents, such as influencers, are able to attract users towards their opinion and make consensus harder to achieve~\citep{chen-2016}. 

This paper studies how interactions outside a pre-defined social network and heterogeneous popularity jointly affect opinion dynamics. We represent these interactions using a multiplex social network, with a complex social network layer and a fully connected explore layer. In the baseline model, interactions on the explore layer allow the mixing of users opinions across the population independently of the social network structure. We then introduce a time-evolving popularity mechanism that biases partner selection on this explore layer, such that frequently selected users become increasingly likely to be re-selected due to changes in their popularity level. The two mechanisms therefore have competing effects. On one hand, the explore layer broadens the set of potential contacts, whereas popularity reinforcement concentrates these interactions on a small number of users. We study how this competition affects the formation of opinion clusters and whether sufficiently concentrated popularity can counteract the homogenising effect of interactions outside the social network.

\section{Methods}\label{sec:meth}
\subsection{Networks}

A social network consists of $N$ nodes, each node $i$ representing a user $i$, and $E$ edges $(i,j)$ representing the social ties between the users. We use a so-called multilayer network which consists of two layers containing the same set of $N$ nodes. The first layer is a social network and represents interactions through social ties. The second layer, which we call the explore layer, defines a fully connected set of potential interaction partners, allowing that every user can interact with any other user independently of the social network structure. At each explore-layer interaction, however, an active user selects only one interaction partner according to the selection rule defined below.

The structure of a social network can be characterised using different measures~\cite{Newman2010}. The degree $k_i$ of user $i$ is the number of social ties to that node. The degree distribution $P(k)$ is the fraction of users in the network with degree $k$, $P(k) = N_k/N$, where $N_k$ is the number of users with degree $k$. The mean degree $\langle k \rangle = \frac{1}{N}\sum_{i=1}^{N} k_i = 2E/N$ measures the average number of social ties in the network. The local clustering coefficient $cc_i$ measures the fraction of pairs of neighbours of users $i$ that are themselves connected, $cc_i = 2e_i/(k_i(k_i-1))$, where $e_i$ is the number of social ties between the neighbours of user $i$. The average clustering coefficient of the network is then defined as $\langle cc \rangle = \frac{1}{N}\sum_{i=1}^{N} cc_i$.

We will use different network models to generate the social network structure. The Erd\H{o}s-R\'enyi (ER) network \citep{Erdos1960} is constructed by considering a set of $N$ users, from which $M$ social ties are chosen uniformly at random. The ER network has a homogeneous degree distribution and a small average clustering coefficient. The Watts-Strogatz (WS) network \citep{watts_collective_1998} is used to generate networks with a larger clustering coefficient. One starts with a cyclic network of users, where each user is connected to 10 nearest neighbours. Social ties are rewired with a probability $q$ to keep high clustering with low distance between users. The configuration (CM) model is used to generate networks with heterogeneous distributions of social ties~\citep{Newman2010}. A degree sequence is generated by drawing the user degree from a power-law distribution, $P(k) \propto k^{-\gamma}$. Each user is assigned a number of stubs corresponding to its degree, after which random pairs of stubs are connected to form social ties. We use $N=1000$ and $\langle k\rangle \approx 10$. Self-loops and multiple social ties between the same pair of users are removed, such that the resulting network is simple and undirected. For each case, we generate 100 samples of the network with the same configuration.

We also use one empirical social network to capture a richer real-life network structure. In the former social media platform Hamsterster, one could create an account for their pets (users) and form friendships between them (social ties)~\citep{nr-aaai15}. This social network is treated as undirected and unweighted. We focus on its largest connected component, consisting of 2000 users, $\langle k\rangle \approx 16$, and $\langle cc\rangle \approx 0.54$. The degree distribution of the network is right skewed. This combination of a large average clustering coefficient and right-skewed degree distribution has been observed in various empirical social networks~\cite{Costa2011SurveyApplications}.

\subsection{Opinion Dynamics Model}

Consider $N \in \mathbb{N}$ users where every user $i$ is specified by their opinion $x_i(t) \in [0,1]$, with $t \in \mathbb{N}$ the discrete time. Initially, each user $i$ draws their opinion $x_i(0)$ from a uniform distribution on the interval $[0,1]$. Every user has a probability $P_a$ of being active at every time step $t$. At each time step $t$, an active user $i$ interacts with a user $j$. If the distance $d$ between their opinions is
\begin{equation}
\label{eq:01}
    d = |x_i(t-1)-x_j(t-1)| \leq \epsilon,
\end{equation}
the active user updates their opinion according to
\begin{equation}
\label{eq:02}
    x_i(t) = x_i(t-1) + \mu\left(x_j(t-1)-x_i(t-1)\right),
\end{equation}
otherwise the opinion remains unchanged. The confidence bound $\epsilon$ captures the open-mindedness of the users, i.e., how susceptible they are to differing opinions. The convergence parameter $\mu \in (0,0.5]$ represents how much a user changes their opinion towards the other user. Only the active user $i$ updates their opinion during each interaction, while the opinion of the selected user $j$ remains unchanged. The interaction is thus directional.

Users can interact with their neighbours in the social network layer, representing interactions with social media friends, or with users in the fully connected explore layer, the latter providing system-wide interactions. In the baseline model, at every time step $t$, we first check if user $i$ is active. An active user $i$ then interacts with one of its social contacts, chosen uniformly at random, with probability $1-P_r$. Otherwise, user $i$ interacts in the explore layer with one user $j$ chosen uniformly at random from all possible users, independently of the social network structure. Since $P_r$ denotes the probability that an active agent interacts in the explore layer, $P_r=0$ indicates that users can only interact on the social network layer, corresponding to a single-layer model. If $d \leq \epsilon$, agent $i$'s opinion is updated according to~(Eq.~\ref{eq:02}).

On social media, some users may be more popular than others and will consequently be seen more often. To include this mechanism in the model, we assign to each user $i$ a popularity score $P_i(t)$, that evolves together with the opinion $x_i(t)$. Initially, all users have the same popularity level $P_i(0)=1$, to control against any initial advantage. The normalised popularity of user $i$ at time $t$ is defined as
\begin{equation}
\label{eq:03}
    p_i(t)=\frac{P_i(t)}{\sum_{l=1}^{N}P_l(t)}.
\end{equation}

The user popularity can increase according to two mechanisms. The first is individual popularity, when user $i$ chooses user $j$ in the explore layer, the popularity of user $j$ simply increases by one, i.e. $P_j(t+1)=P_j(t)+1$. This mechanism captures the growth of popularity through social interactions. The second is via viral events, capturing the occasional large increase in the popularity of a user due to a particular post or event. At every $f_v$ time steps one user $j$ is randomly chosen with probability proportional to their current popularity. This viral event increases the popularity of user $j$ by $i_v$. We refer to $f_v$ as the viral-event interval and $i_v$ as the viral impact. This positive feedback is closely related to a Pólya-type reinforcement~\cite{Pemantle2007} where a user is selected according to its current popularity, and being selected further increases that popularity in follow up steps.

We thus define two complementary popularity-driven models. The individual model assumes that users only increase their popularity through the individual mechanism, whereas the viral model assumes that both mechanisms may affect the user popularity. In both models, an active user interacting in the explore layer selects user $j$ with probability $p_j(t)$, rather than uniformly as in the baseline model. The opinion update then follows~(Eq.~\ref{eq:02}). The difference with the baseline model is in the selection of the interaction partner on the explore layer. Instead of choosing an agent uniformly at random, agent $i$ chooses an agent $j$ with a probability proportional to agent $j$'s normalised popularity $p_j(t)$.

For each model, we run the simulations for 100 versions of the social network using the same set of parameters. For the empirical network, we run the simulations 100 times. We fix $\mu=0.5$ and $P_a=0.5$. In the popularity-driven models, we first let the popularity dynamics evolve independently and after a transient of 500 time steps, we allow the popularity and the opinions to update via the interactions. That means, at the start of the opinion dynamics the agents have already generated a heterogeneous popularity. The dynamics evolve until all the opinions have converged, i.e. the user's opinion has not changed by more than $\max_i |x_i(t)-x_i(t-1)| < 0.005$ during $\Delta T=600$ consecutive time steps.

\begin{figure}
\centering
\includegraphics[width=0.45\textwidth]{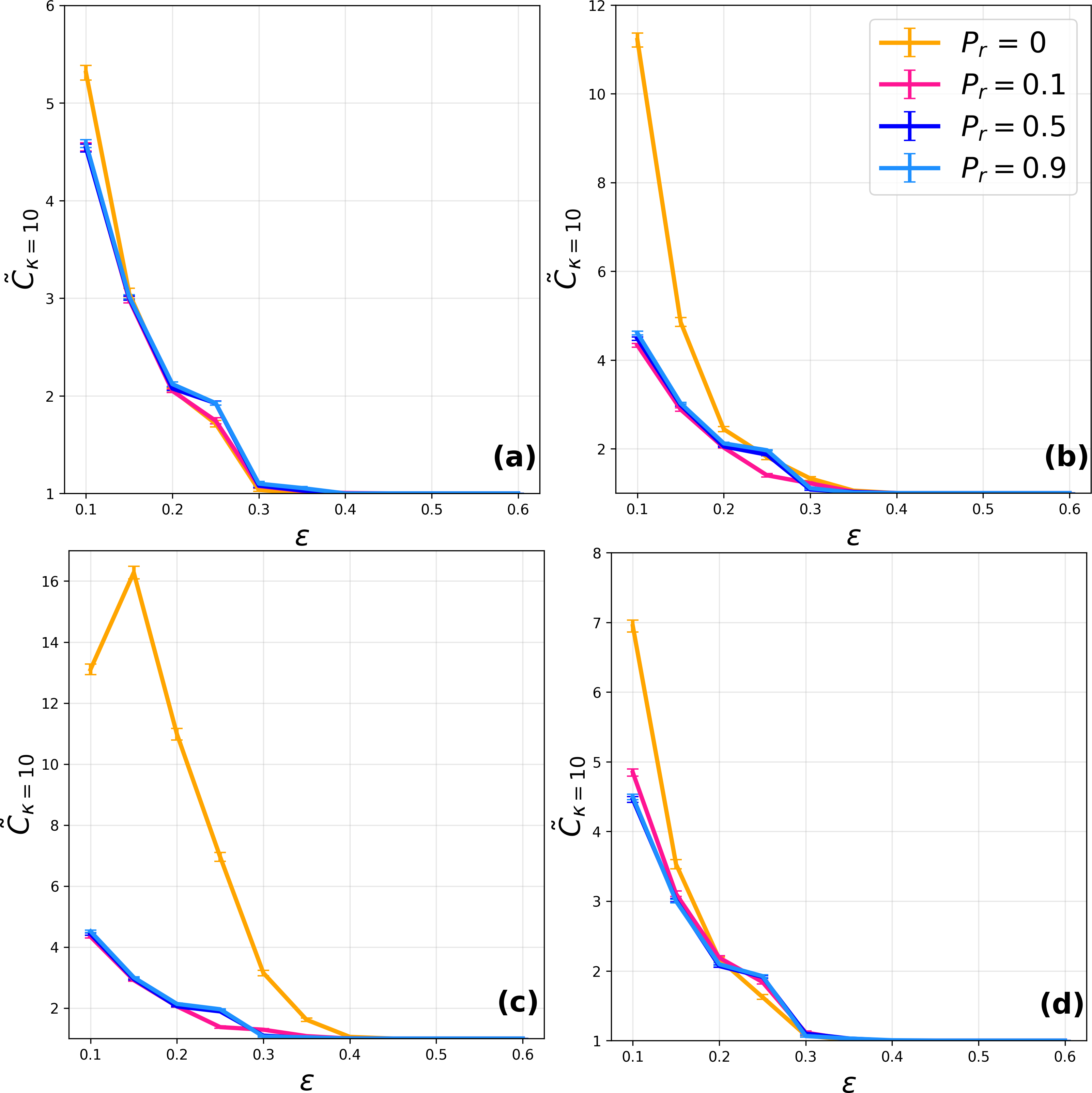}
\caption{Number of major opinion clusters $\tilde{C}_{\kappa=10}$ as a function of the confidence bound $\epsilon$ for the baseline model with the (a) ER, (b) WS01 (WS with $q=0.1$), (c) WS001 (WS with $q=0.01$), and (d) CM social networks.}
\label{fig1}
\end{figure}

The opinion interval $[0,1]$ is divided into $100$ equally sized bins. We denote the $l$-th opinion cluster containing at least $\kappa$ users by $\tilde{C}_{\kappa,l}$. A cluster is defined as a connected set of non-empty bins separated from other clusters by bins with zero mass. The total number of clusters containing at least $\kappa$ users is denoted by $\tilde{C}_{\kappa}$. Clusters containing at least $\kappa=10$ users are called major clusters. Let $m(\tilde{C}_{\kappa,l})$ denote the number of users in cluster $l$. The effective number of clusters is defined by weighting each cluster $l$ by its size $m(\tilde{C}_{\kappa=1,l})$ (Eq.~\ref{eq:04}). The sum runs over all clusters, such that larger clusters contribute more strongly to the measure. We use opinion diversity or fragmentation to describe states with multiple opinion clusters, while polarisation refers specifically to states characterised by two separated major opinion clusters.

\begin{equation}
\label{eq:04}
\tilde{C}_{\mathrm{eff}} = \frac{ \left(\sum_{l} m\!\left(\tilde{C}_{\kappa=1,l}\right)\right)^2}{\sum_{l} \left[m\!\left(\tilde{C}_{\kappa=1,l}\right)\right]^2}
\end{equation}

\begin{figure*}[!t]
\centering
\includegraphics[height=0.45\textheight]{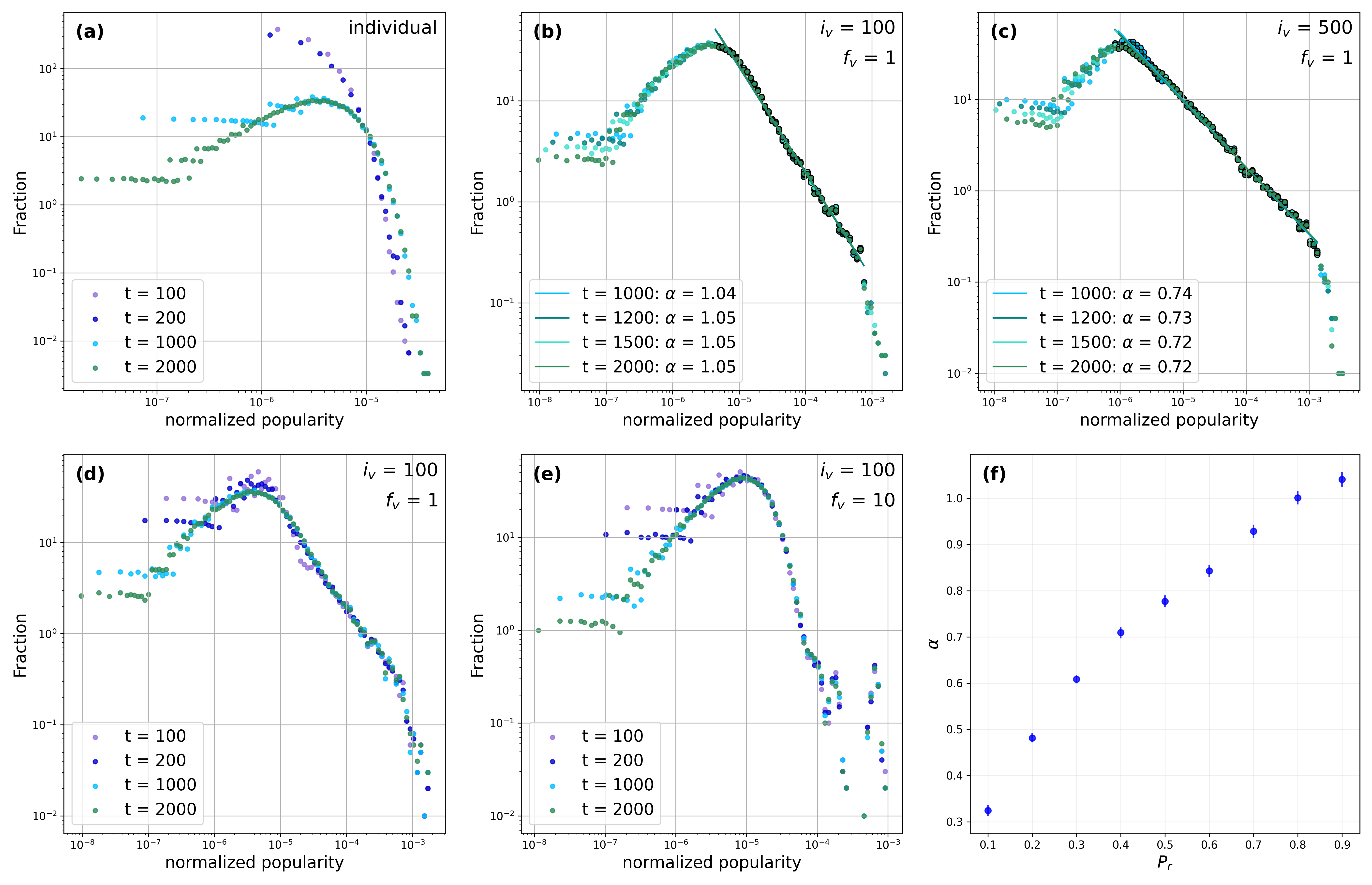}
\caption{Popularity distributions and their dependence on the model parameters. (a) Popularity distribution in the individual model with $P_r = 0.9$; (b) Popularity distribution of a viral model with $P_r = 0.9$, $i_v = 100$ and $f_v = 1$; (c) Popularity distribution of a viral model with $P_r = 0.9$, $i_v = 500$ and $f_v = 1$; (d) Popularity distribution of a viral model with $P_r = 0.9$, $i_v = 100$ and $f_v = 1$; (e) Popularity distribution of a viral model with $P_r = 0.9$, $i_v = 100$ and $f_v = 10$; (f) Relation between the fitted slope (using a least-squares fit on the best fitting window) of the scale-free interval of the popularity distribution and $P_r$, with $i_v = 100$ and $f_v = 1$.}
\label{fig2}
\end{figure*}

\section{Results}\label{sec:resu}

\subsection{Multilayer Opinion Dynamics}

Figure~\ref{fig1} shows that independently of the social network structure, coupling the social network to the explore layer reduces the number of major opinion clusters at small confidence bounds $\epsilon$. This reduction is stronger and persists to larger values of $\epsilon$ for networks with higher clustering. Most of the reduction occurs already at small mixing (small $P_r$), while further increasing $P_r$ produces only minor changes in the number of major clusters. Compared with the ER network, the structurally heterogeneous CM network and particularly the highly clustered WS networks sustain greater opinion diversity at small confidence bounds. Yet, even a small $P_r$ strongly reduces the differences between network structures due to the coupling with the explore layer increasing the diversity in interacting agents. These results indicate that uniformly sampled interactions on the explore layer act as a global mixing mechanism. They connect users beyond their direct social neighbourhoods, making the opinion dynamics less dependent on a particular network structure, which leads to a reduction of opinion fragmentation.

\subsection{Emergence of Popularity Heterogeneity}

In the individual model, there are no viral events and the popularity of the users depends entirely on being chosen by other users in the explore layer. This process does not affect all agents equally because popularity reinforces the chance of future selection. Users who are chosen early on and increase their popularity in the first time steps have a higher chance of being chosen in subsequent time steps, leading to a further increase in popularity due to reinforcement. Figure~\ref{fig2}a shows that independently of the time step $t$, the emergent distribution is not skewed and most of the users have low to intermediate popularity levels.

In contrast, the quasi-stationary popularity distribution for viral models is right-skewed if the viral impact $i_v$ is large enough (Fig.~\ref{fig2}b-e). Viral models produce some users with a significantly higher popularity than the rest. If $i_v$ is small, there are no high-popularity users and the popularity distribution resembles that of the individual model. A larger viral impact, together with the reinforcing process, results in a flatter popularity distribution compared with smaller values of the viral impact, for which users experiencing a viral event receive less reinforcement. Figure~\ref{fig2}f shows that the fitted slope increases when the probability of interacting in the explore layer is increased, due to the larger contribution of the individual popularity mechanism in the viral model. Furthermore, increasing $f_v$ results in a decrease in popularity heterogeneity due to the reduced contribution of viral events to popularity (Figs.~\ref{fig2}d,e).

\subsection{Popularity and Opinion Dynamics}

\begin{figure}[]
    \centering
    \includegraphics[width=\linewidth]{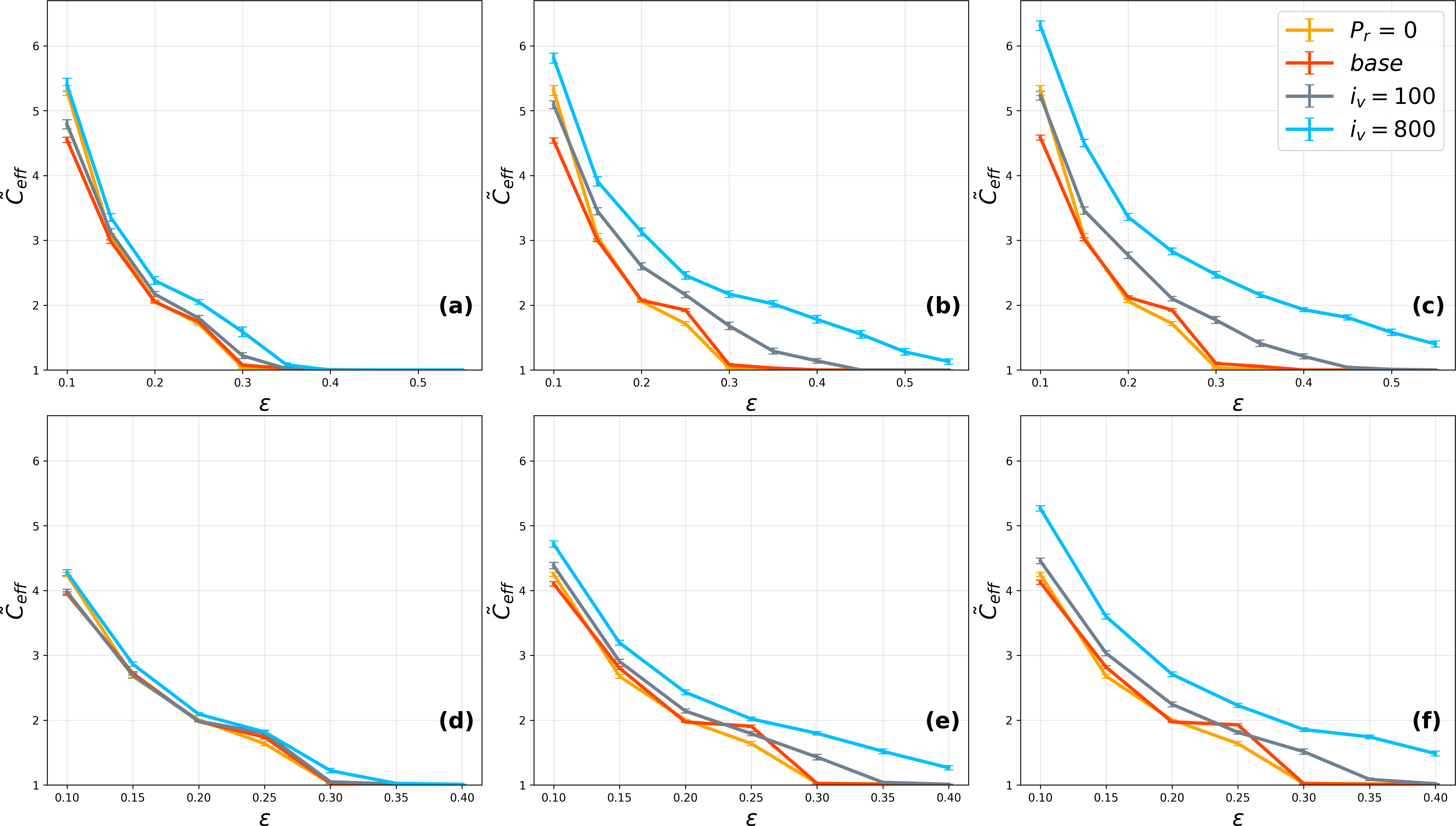}
    \caption{Opinion clusters for different viral models.
    Panels (a--c) show the number of major clusters
    $\tilde{C}_{\kappa=10}$, while panels (d--f) show
    the effective number of clusters $C_{\mathrm{eff}}$.
    Panels (a,d), (b,e), and (c,f) correspond to
    $P_r=0.1$, $0.5$, and $0.9$, respectively.
    We use a random ER network and $f_v=1$.}
    \label{fig3}
\end{figure}

Figure~\ref{fig3} shows that the introduction of the viral mechanism can increase the opinion diversity due to the emergence of high-popularity users. For all three values of $P_r$, increasing the viral impact $i_v$ can increase opinion diversity relative to the baseline model. As $i_v$ increases, interactions on the explore layer become increasingly concentrated on a small number of highly popular users. This concentration balances the homogenising effect of uniform explore-layer interactions, sustaining additional opinion clusters. The effect increases with $i_v$ until a small subset of users contains most of the popularity, after which a further increase in $i_v$ has little additional effect.

For $P_r = 0.1$, the results remain close to those obtained for the baseline model. Even though a small fraction of users accumulates most of the total popularity for $i_v \geq 100$, thus dominating interactions in the explore layer, the low probability of explore-layer interactions limits their overall effect for all values of $i_v$. For small $\epsilon$, large $i_v$ increases opinion diversity, making the viral model more similar to the single-layer case. For $P_r = 0.1$, the corresponding differences in $\tilde{C}_{eff}$ is negligible.

If the coupling between both layers increases ($P_r = 0.5$), the number of interactions with high-popularity users also increases. For both $i_v = 100$ and $i_v = 800$, the introduction of viral events increases opinion diversity for all $\epsilon$ compared with the baseline model. For $\epsilon = 0.1$, only models with large $i_v$ result in more major clusters than the single-layer model. For larger values of $\epsilon$ and $i_v \geq 100$, viral events increase opinion diversity and can sustain polarisation in regimes where the single-layer and baseline models converge to consensus. When two major clusters are present, $\tilde{C}_{eff}$ indicates that the two clusters are of similar size. For $\epsilon = 0.4$--$0.5$, viral models with large popularity heterogeneity can still converge to a polarised state, but with a lower probability and a larger difference in cluster sizes. For $P_r = 0.9$, these effects become more pronounced: models with the highest popularity heterogeneity have one more major opinion cluster than the single-layer model up to $\epsilon \approx 0.4$ and sustain polarisation over a wider range of $\epsilon$ than the baseline model. At small $\epsilon$ the main effect of popularity is an increase in fragmentation, whereas at larger $\epsilon$, where the baseline and single-layer models approach consensus, strong popularity heterogeneity can instead sustain a two-cluster polarised state.

\begin{figure}[]
\centering
\includegraphics[width=\linewidth]{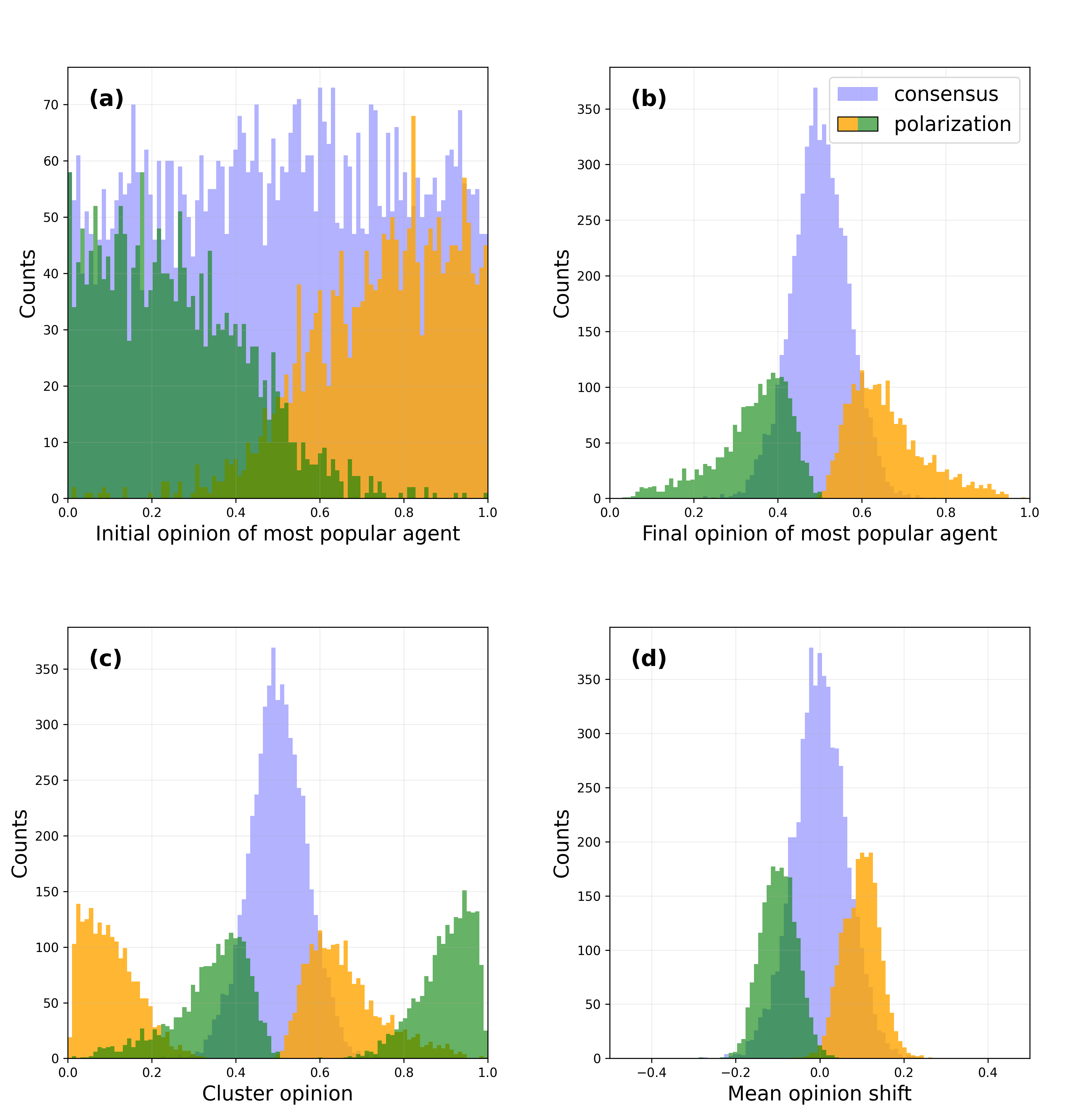}
\caption{Influence of the most popular user on the opinion dynamics. (a) initial opinion of the most popular user; (b) final opinion of the most popular user; (c) opinion of the final clusters; (d) shift of the mean opinion. Orange indicates that the opinion of the most popular user is larger than $0.5$, while green indicates that it is smaller than $0.5$. Measures obtained over 10000 realisations of the viral model on an ER network with $N=1000$, $P_r=0.5$, $P_a=0.5$, $f_v=1$, $i_v=800$, and $\epsilon=0.45$.}
\label{fig4}
\end{figure}

Figure~\ref{fig4} illustrates this polarisation mechanism for $P_r=0.5$, $i_v=800$, and $\epsilon=0.45$. The increase in opinion diversity can be attributed to highly popular users acting as opinion attractors, pulling other users towards their opinions through interactions in the explore layer (Fig.~\ref{fig4}b,c). Users that are not drawn towards the most popular user, either because their opinion difference exceeds the confidence bound or because they are influenced by another highly popular user, can form a separate cluster. This attractor mechanism is strengthened by the asymmetric interaction rule. When a highly popular user is selected as an interaction partner, it influences the active user without changing its own opinion. Highly popular users can still change their opinions when they are themselves active and select another user. This directional influence can shift the mean opinion and allow well-separated opinion clusters to persist (Fig.~\ref{fig4}d). These influential users are not introduced as fixed opinion leaders. In fact, their influence emerges from the popularity dynamics because they become increasingly likely to be selected as interaction partners.

\begin{figure}[]
\centering
\includegraphics[width=0.28\textwidth]{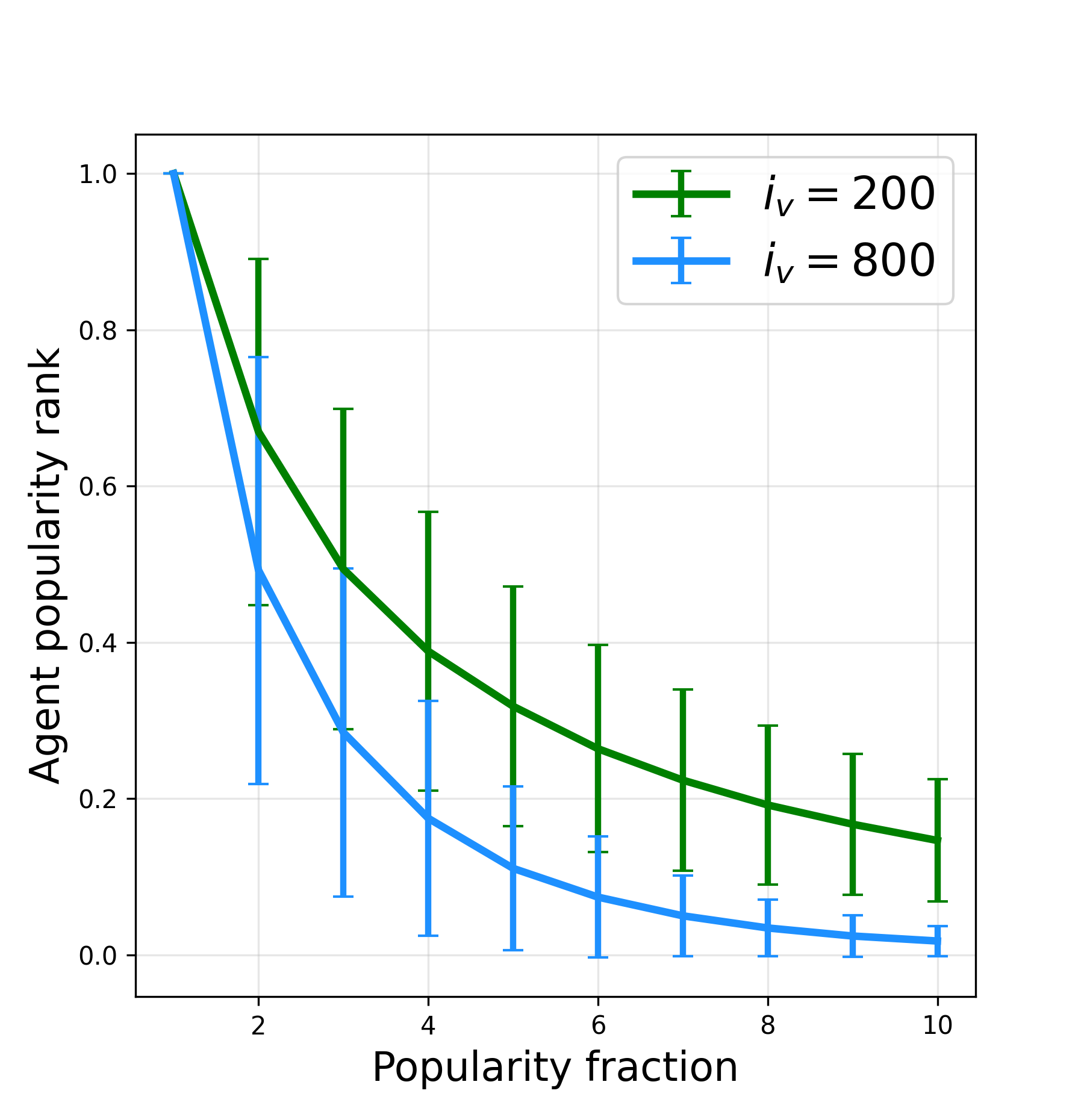}
\caption{Fraction of the popularity of the 10 most popular users relative to the most popular user for viral models with $N=1000$, $P_a=0.5$, $P_r=0.5$, $f_v=1$, and $i_v=200$ or $i_v=800$. Error bars indicate the standard deviation across $1000$ realisations.}
\label{fig5}
\end{figure}

Although the popularity distributions for $i_v = 200$ and $i_v = 800$ appear similar, the logarithmic binning can obscure differences among the most popular users. Figure~\ref{fig5} shows that for $i_v = 800$, popularity decreases more rapidly with rank relative to the most popular user. Thus, interactions in the explore layer are more strongly concentrated on the most popular user, strengthening its role as an opinion attractor and helping to explain the larger number of major clusters.

\begin{figure*}
\centering
\includegraphics[width=0.8\linewidth]{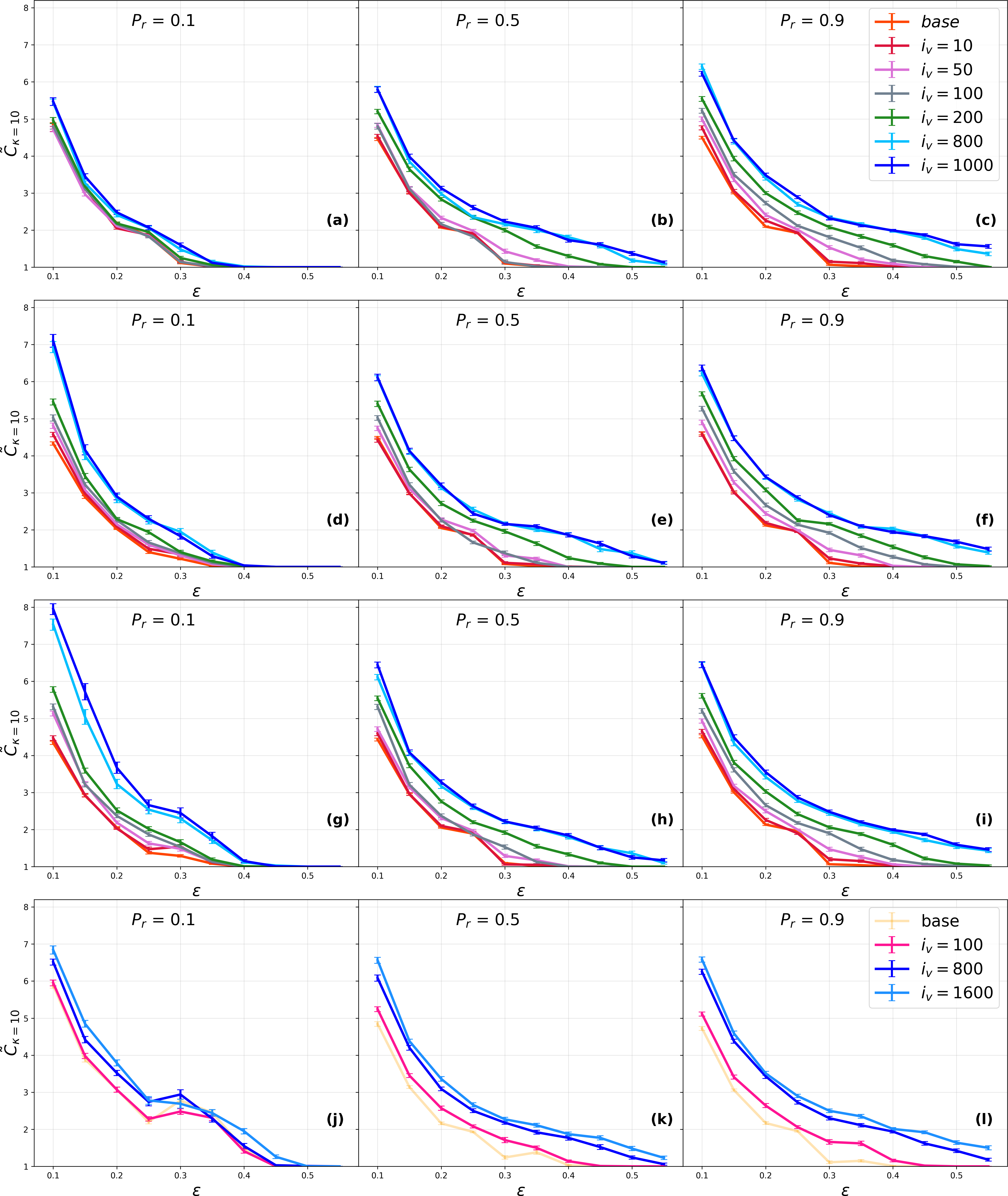}
\caption{Number of major opinion clusters $\tilde{C}_{\kappa=10}$ as a function of $\epsilon$ for viral models with $f_v=1$ and varying $i_v$: (a-c) CM; (d-f) WS with $q=0.1$; (g-i) WS with $q=0.01$; and (j-l) empirical network. The columns correspond to $P_r=0.1$, $0.5$, and $0.9$.}\label{fig6}
\end{figure*}

\subsection{Network Structure and Opinion Dynamics}

We now analyse how the popularity-driven mechanism depends on the underlying social network structure. The viral dynamics on a CM structure (Fig.~\ref{fig6}a-c) is similar to the ER case (Fig.~\ref{fig3}). For $P_r = 0.1$ the viral model can increase the number of major opinion clusters for all values of $\epsilon$. For larger values of $P_r$, the results also resemble what was observed for other network structures, indicating that frequent interactions in the explore layer weaken the influence of the underlying social network topology.

A different behaviour appears for the highly clustered WS networks. For $P_r = 0.1$, small $\epsilon$, and large $i_v$, the WS01 network produces more major clusters than for larger values of $P_r$ (Fig.~\ref{fig6}d-f). In contrast, for the ER network, the number of major opinion clusters increases with $P_r$ at $\epsilon = 0.1$ and large $i_v$. This difference results from the interplay between the two network layers. When $i_v$ is large, a small number of users accumulate most of the popularity and dominate interactions in the explore layer. At small confidence bounds, however, many of these interactions do not lead to opinion updates because the opinion difference exceeds $\epsilon$. Since $P_r$ is small, the opinion dynamics therefore remains largely governed by interactions on the social network layer. The high clustering coefficient of the WS01 network promotes greater opinion diversity. Despite this increase, the viral model still produces fewer major clusters than the corresponding single-layer model up to $\epsilon \approx 0.3$. For larger $\epsilon$, however, the viral model can sustain a polarised state while the single-layer and baseline models converge to consensus. The increase in the number of clusters for small $P_r$, small $\epsilon$, and large $i_v$ is stronger for the WS001 network, consistent with the larger clustering coefficient (Fig.~\ref{fig6}g-i). For $P_r \in \{0.5,0.9\}$, the viral dynamics show much weaker dependence on the underlying structure. Because the corresponding single-layer dynamics remain strongly dependent on network structure, the viral model can produce either more or fewer opinion clusters than the single-layer model depending on the structure and the value of $\epsilon$. Figure~\ref{fig6}j-l shows that the empirical network follows the same qualitative behaviour. Increasing the viral impact $i_v$ increases opinion diversity for all values of $P_r$. At small $P_r$, the outcome remains more strongly influenced by the underlying network structure, whereas at larger $P_r$, the dynamics is increasingly dominated by interactions with highly popular users.

The results show that the relative importance of the social network structure and popularity-driven interactions depends on both $P_r$ and $i_v$. At small $P_r$, the network structure remains important because most interactions still occur through the social layer, while increasing $i_v$ leads to an increase in the popularity heterogeneity generated by viral events. At large $P_r$, interactions in the explore layer become more frequent, so the effect of increasing $i_v$ becomes more pronounced, i.e.\, a small number of highly popular users increasingly dominate the dynamics and can sustain polarisation over a wider range of $\epsilon$.


\section{Conclusions}\label{sec:conc}

Individuals change their opinion through repeated social interactions, which are constrained by social networks. On social media, users are also exposed to content outside their immediate social neighbourhood, and a small number of highly visible users may be encountered much more often than the majority of others. In this paper, we studied how the coupling of a social network layer to an explorative interaction layer, and the introduction of popularity-driven interactions on this second layer, affect the formation of opinion clusters in a bounded-confidence opinion dynamics model.

Adding an explore layer to an underlying social network increases the diversity of interaction partners and reduces opinion diversity. In the single-layer model, highly clustered networks produce more opinion clusters at small confidence bounds. Explore-layer interactions strongly suppress this additional fragmentation, reducing the differences between network structures. Consequently, opinion diversity becomes less dependent on the underlying social network structure as users interact more frequently through the explore layer.

The popularity heterogeneity generated by the individual popularity mechanism is not large enough to substantially affect the final number of major opinion clusters relative to the baseline model. In contrast, viral events create popularity heterogeneity, which can increase opinion diversity compared to the baseline model. Highly popular users act as local attractors, pulling other users' opinions towards them and leading to the formation of additional opinion clusters. For sufficiently large popularity heterogeneity, this mechanism can sustain polarisation in regimes where the corresponding model without popularity converges to consensus. In other words, popularity heterogeneity can counteract the reduction in opinion diversity caused by the increased interaction diversity of the explore layer.

These results should be interpreted with caution with respect to empirical social media. Our model is simplified, as opinions are represented by a single continuous variable, interactions are pairwise, users are homogeneous except for popularity, and interactions on the explore layer are based only on popularity. Future work could include both popularity and homophily in the selection of interaction partners, to capture more realistic interaction dynamics. More heterogeneous users could also be considered, for example by allowing the confidence bound or convergence parameter to vary between users or depend on popularity.

In conclusion, the emergence of opinion clusters in social media-inspired networks depends not only on the underlying network structure, but also on the diversity and heterogeneity of interactions. The explore layer tends to reduce opinion diversity by increasing the diversity of interaction partners, whereas strong popularity heterogeneity can reduce this effect by creating highly influential users acting as attractors. Consequently, a small number of highly popular users can qualitatively change the opinion dynamics and sustain polarisation in regimes where the corresponding model without popularity would converge to consensus. These results suggest that popularity amplification may provide an additional control parameter for collective opinion dynamics, alongside network structure and confidence bounds. Future research could study how modifying the concentration of exposure affects transitions between consensus, fragmentation, and polarisation.

\section{Acknowledgements}\label{sec:ack}

L.E.C.R. was supported by the Belgian Special Research Fund (BOF) (2024/01/709).

\printcredits

\bibliographystyle{unsrtnat}
\bibliography{ref_paper}



\end{document}